# Conversational Swarm Intelligence, a Pilot Study

**Louis Rosenberg**
Unanimous AI
Pismo Beach, CA
louis@unanimous.ai

**Gregg Willcox**
Unanimous AI
Seattle, Washington
gregg@unanimous.ai

**Hans Schumann**
Unanimous AI
San Francisco, CA
hans@unanimous.ai

**Miles Bader**
Vassar College
Poughkeepsie, NY
mbader@vassar.edu

**Ganesh Mani**
Carnegie Mellon University
Pittsburgh, PA
ganeshm@andrew.cmu.edu

**Kokoro Sagae, Devang Acharya, Yuxin Zheng, Andrew Kim, and Jialing Deng**
Carnegie Mellon University

**ABSTRACT**

Conversational Swarm Intelligence (CSI) is a new method for enabling large human groups to hold real-time networked conversations using a technique modeled on the dynamics of biological swarms. Through the novel use of conversational agents powered by Large Language Models (LLMs), the CSI structure simultaneously enables underline{local dialog} among small deliberative groups and underline{global propagation} of conversational content across a larger population. In this way, CSI combines the benefits of small-group deliberative reasoning and large-scale collective intelligence. In this pilot study, participants deliberating in conversational swarms (via text chat) (a) produced 30% more contributions (p<0.05) than participants deliberating in a standard centralized chat room and (b) demonstrated 7.2% less variance in contribution quantity. These results indicate that users contributed underline{more content} and participated underline{more evenly} when using the CSI structure.

**CSS Concepts**
•Human-centered computing | Human computer interaction (HCI) | Collaborative and social computing

**INTRODUCTION**

Collective Intelligence (CI) refers to the field of research in which the knowledge, wisdom, and insights of human groups is collected and processed to achieve more accurate insights than individuals could produce on their own [1]. Common methods for tapping the intelligence of human groups are based largely on votes, polls, and surveys of various forms. These methods have been modernized in recent years through online upvoting systems, prediction markets, and sentiment extraction via language processing (NLP), but still rely heavily on collecting isolated responses from individual participants and aggregating those responses into statistical profiles that often mask important beliefs or attitudes.

An alternate method called **Artificial Swarm Intelligence** (ASI) has been tested in recent years and found to outperform common statistical methods, especially when populations harbor diverse or conflicting views [2-7]. It works by enabling networked human groups to form real-time systems (i.e., *swarms*) that interact and converge on unified solutions. The technology of ASI was first proposed in 2015 and uses algorithms modeled on the emergent behaviors of schooling fish, flocking birds, and swarming bees [2,3]. Biologists refer to these emergent behaviors as Swarm Intelligence because it turns biological groups into "super-organisms" that can make more effective decisions than the individuals can make on their own [1, 8, 9]. ASI enables human groups to achieve similar benefits, amplifying group intelligence through real-time organic interactions [10-11].

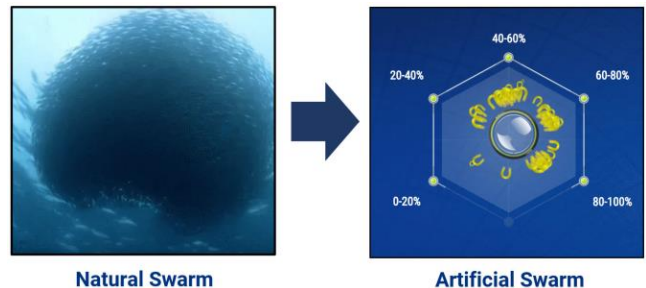

**Figure 1: Natural Swarm vs ASI system of 100 participants.**

As shown in Figure 1 above, current methods for Artificial Swarm Intelligence involve use-cases where human groups deliberate among predefined sets of options and collectively rate, rank, or select among them. For example, swarm can be used for (a) underline{forecasting} the most likely outcome from a set of possible outcomes, (b) underline{prioritizing} sets of options into ranked lists that optimizes group satisfaction, or (c) underline{rating} the relative strengths of various options against specific metrics [4-6]. While these capabilities are useful in many real-world applications, there is a need for a more flexible methods of deploying ASI, especially for enabling groups to deliberate open-ended questions in which a set of potential solutions to select among are underline{not} known in advance.

In the following sections, we will describe a new approach for connecting networked human participants into real-time ASI systems, leveraging the traditional benefits of Swarm Intelligence through flexible conversational interactions.

## CONVERSATIONAL SWARM INTELLIGENCE (CSI)

To address the limitations of prior CI systems modeled on the principles of Swarm Intelligence, a new architecture called Conversational Swarm Intelligence (CSI) has been developed, deployed, and piloted in an initial study. The motivation for the CSI architecture is to enable large human groups to deliberate conversationally as real-time dynamic systems (i.e., swarms) that converge organically on solutions that maximize group satisfaction. By enabling the swarming process to occur conversationally, we can eliminate the need for predefined options. Instead, options emerge organically as participants suggest and debate ideas in real-time.

Current ASI systems typically engage human groups ranging from 20 to 250 real-time participants. The goal of CSI is to enable conversational deliberation among similar sized groups or larger. This poses a unique challenge for an online conversational system. For example, bringing 100 people into a single chatroom would not yield meaningful dialog or insight. That's because conversational quality degrades with group size [12]. Sometimes referred to as the "many minds problem," when groups grow beyond a handful of people, the conversational dynamics fall apart, providing less "airtime" per person, disrupting turn-taking dynamics, providing less feedback per comment, and reducing engagement as participants feel less social pressure to participate.

Other research suggests that the ideal group size for real-time deliberation ranges from 4 to 7 members, with conversations transitioning from authentic dialog to sequential monologue as groups approach 10 members [16]. Other research suggests that maximum satisfaction for participants occurs around 5 members [17]. For these reasons, putting 100 people in a standard chatroom does not generally yield authentic deliberative "conversations" but merely a stream of singular comments with little interaction among them.

Fish schools, on the other hand, can hold "conversations" among hundreds or thousands of members with no central authority mediating the process. Each fish communicates with others using a unique organ called a "lateral line" that senses pressure changes caused by neighboring fish as they adjust speed and direction with varying levels of conviction. The number of neighbors that a given fish pays attention to varies from species to species, but it's always a small subset of the group. And because each fish reacts to an overlapping subset of other fish, information quickly propagates across the full population, enabling a single Swarm Intelligence to emerge that rapidly converges on unified decisions [13]. This is a powerful biological solution that was first emulated in networked human groups in 2021 using a technology called "hyperswarms" and was shown to enable information propagation across overlapping groups [14].

Researchers at Unanimous AI have recently applied this method to real-time human conversations. We call this a "HyperChat" structure and it's modeled on schooling fish. In this method, a group of 250 online participants could be broken up into a large number of smaller groups, for example 50 groups of 5 people, with each group put into their own chat room and asked to discuss an issue in parallel with the other groups. This alone does not yield Swarm Intelligence because information cannot propagate across the population. We solved this using AI Agents powered by Large Language Models (LLMs) to emulate the function of the lateral line organ in fish. In particular, we insert an **AI Agent** into each of the 50 chat rooms that monitors the dialog in that room, distills the salient content, and expresses that content in a neighboring room through natural first-person dialog. In this way, each of the 50 groups is given a 6th member that happens to be an AI Agent whose function is to express (at intervals) the insights that emerge in one group into neighboring groups, thereby enabling information to propagate across the full "swarm."

This creates a single real-time system in which 250 or 2500 or even 25,000 people could hold a conversation on a single topic and converge on unified solutions that optimize groupwise support. In this way, we enable large populations to interact conversationally in real-time while ensuring that (a) individual members can have meaningful discourse in small deliberative groups, and (b) information propagates globally in near real-time, leveraging the benefits of inactive swarms. An example structure for CSI is shown in Figure 2.

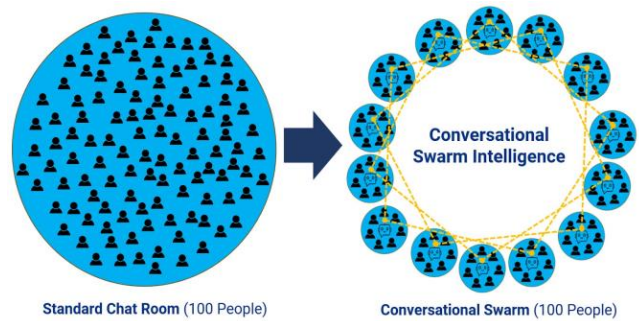

**Figure 2: Standard Chat versus a Conversational Swarm**

This method has two significant benefits over prior ASI systems. First, it enables the use of open-ended questions, enabling participants to suggest and debate options that are not pre-defined. And second, it allows users to not only indicate which options they prefer but also freely discuss why they prefer them. In this way, the CSI methodology not only elicits conversational solutions that maximize collective support but also captures the reasons why the group supports (or rejects) the various options raised. Also, the structure in Figure 2 reduces the Social Influence Bias that hinders many real-time CI methods such as online discussion forums, chat rooms, and sequential upvoting systems.

For example, when a single user posts an idea, comment, or criticism into a standard forum or chat room, they can bias the full population [15]. This means that comments posted early have an overweight effect in sequential systems. It also means that a small number of strong personalities can have an overweight effect in parallel systems. We expect CSI to mitigate social influence bias using the unique structure shown above because (a) each individual is only influenced by a small number of others in real-time, and (b) ideas only propagate organically after they gain local momentum.

## PILOT STUDY

A pilot study was run to evaluate the effectiveness of conversational swarms in forming consensus judgements in groups of roughly 25 users. A total of 10 questions were asked to the same group, half using a standard chat room and half using a conversational swarm. In each case, four questions asked the participants to choose the best next move in a chess puzzle, while one question asked an open-ended AI ethics question. Each question was limited to 5 minutes or less of discussion, after which a final answer was reported.

For the trials that used Conversational Swarm Intelligence, the ~25 users were divided into 5 subgroups of ~5 users, each subgroup connected to a neighboring subgroup by an AI agent powered by GPT-3.5. Each AI agent was tasked with observing conversations in one subgroup and passing conversational summaries to another subgroup at one-minute intervals, thereby enabling information propagation.

As shown in Table 1, participants in the CSI structure produced 30% more contributions ($p<0.05$) compared to the standard chat room, measured in the number of suggestions, explanations, agreements, disagreements, and questions made by the group. The variance in contribution also decreased by 7.2%, indicating that users contributed more evenly in the CSI structure than a standard chat room.

| Contribution Quantity | Standard Chat Room | Conversational Swarm |
|---|---|---|
| Mean | 7.28 | 9.48 |
| Variance | 94.6 | 87.8 |

**Table 1. Comparison between Average User Contribution in Conversational Swarms and Standard Chat Rooms.**

## CONCLUSIONS

We introduce Conversational Swarm Intelligence (CSI), a novel technology that allows large, networked groups to hold coherent real-time discussions via online chat and quickly reach groupwise consensus. We performed a pilot study and demonstrated feasibility. Specifically, we collected pilot data showing that using "conversational swarms" results in 30% higher levels of user contribution and 7% less variance in user contribution. We attribute these benefits to the effect of individuals interacting in smaller groups while still being informationally connected to the larger population through the novel use of conversational AI agents. We expect these benefits to increase as we test with larger populations of real-time participants, which future work will validate.

## ACKNOWLEDGEMENTS

We would like to thank the students in Francesca Xhakaj and Ganesh Mani's summer ELAIDA (Experiential Learning from AI and DAta) class for volunteering for the pilot study.